\documentclass[%
 reprint,
 amsmath,amssymb,
 aps,
aps,prd,twocolumn,superscriptaddress
]{revtex4-2}

\usepackage{graphicx}
\usepackage{amsmath}
\usepackage{xcolor}
\usepackage{url}

\begin{document}

\title{Classifying Core-Collapse Supernova Gravitational Waves using Supervised Contrastive Learning}

\author{Ao-Bo Wang}
\affiliation{School of Physics Science And Technology, Wuhan University, No.299 Bayi Road, Wuhan, Hubei, China}

\author{Yong Yuan}
\email{yuanyong@imech.ac.cn}
\affiliation{Center for Gravitational Wave Experiment, National Microgravity Laboratory, Institute of Mechanics, Chinese Academy of Sciences, Beijing, China}

\author{Hao Cai}
\email{hcai@whu.edu.cn}
\affiliation{School of Physics Science And Technology, Wuhan University, No.299 Bayi Road, Wuhan, Hubei, China}

\author{Xi-Long Fan}
\email{xilong.fan@whu.edu.cn}
\affiliation{School of Physics Science And Technology, Wuhan University, No.299 Bayi Road, Wuhan, Hubei, China}

\date{\today}

\begin{abstract}
The detection and reconstruction of gravitational waves from core-collapse supernovae (CCSN) present significant challenges due to the highly stochastic nature of the signals and the complexity of detector noise. In this work, we introduce a deep learning framework utilizing a ResNet-50 encoder pre-trained via supervised contrastive learning to classify CCSN signals and distinguish them from instrumental noise artifacts. Our approach explicitly optimizes the feature space to maximize intra-class compactness and inter-class separability. Using a simulated four-detector network (LIGO Hanford, LIGO Livingston, Virgo, and KAGRA) and realistic datasets injecting magnetorotational and neutrino-driven waveforms, we demonstrate that the contrastive learning paradigm establishes a superior metric structure within the embedding space, significantly enhancing detection efficiency. At a false positive rate of $10^{-4}$, our method achieves a true positive rate (TPR) of nearly $100\%$ for both rotational and neutrino-driven signals within a distance range of $10$--$200$~kpc, while maintaining a TPR of approximately $80\%$ at $1200$~kpc. In contrast, traditional end-to-end methods yield a TPR below $20\%$ for rotational signals at distances $\geq 200$~kpc, and fail to exceed $60\%$ for neutrino-driven signals even at a close proximity of $10$~kpc.
\end{abstract}

\maketitle

\section{Introduction}

In 2015, the Advanced LIGO (aLIGO; \cite{LIGO_2015CQGra}) made the first direct detection of gravitational waves (GWs) from the merger of two black holes, marking the beginning of the era of GW astronomy \citep{Abbott2016PhRvL}. Subsequently, the Advanced Virgo detector (aVirgo; \cite{Acernese_2015}) and KAGRA \citep{Abbott2022PTEP} joined the global network of ground-based GW observatories. By the end of the O4a observing run, the network had detected more than 200 GW events originating from mergers of compact binary systems \citep{LIGO2025arXiv3, LIGO2025arXiv2, LIGO2025arXiv1}. With the continued upgrades of ground-based detectors and the planned construction of the high-frequency detector NEMO \citep{Ackley2020PASA}, as well as the development of next-generation observatories such as the Einstein Telescope \citep{Punturo_2010CQGra} and Cosmic Explorer \citep{Abbott2017CQGra}, an increasing number and greater diversity of GW events and sources are expected to be detected in the near future \citep{Abbott2017CQGra, Maggiore2020JCAP}.

In recent decades, with the rapid progress in multidimensional simulations of core-collapse supernovae (CCSNe), our understanding of the physical mechanisms responsible for GW emission during supernova explosions has significantly deepened (see \cite{Janka_2012, Janka2016ARNPS, Muller2016PASA, Burrows2021Natur} for recent reviews). These simulations have revealed that the frequencies of GWs generated by CCSNe lie within the sensitive bands of the aLIGO, aVirgo and KAGRA detectors \citep{Kotake2006RPPh, Moenchmeyer1991A&A, Muller2011hnse, Janka2007PhR}. Moreover, recent studies suggest that the GW memory signals produced by nearby (kpc-scale) CCSNe could be detectable by space-based detectors such as LISA and Taiji\citep{amaro2017laser,hu2017taiji}, as well as future lunar-based GW observatories \citep{Richardson2024PhRvL, Chio2024ApJ, Richardson2025arXiv}. GWs emitted from the core during the collapse provide a direct probe of the explosion mechanism, enabling us to gain valuable insights into the dynamics of matter motion in the supernova engine \citep{Szczepanczyk2021PhRvD}. However, to date, searches for GW signals from CCSNe have not yielded any significant detection candidates \citep{Abadie2012PhRvD, Abbott2016PhRvD, Abbott2017PhRvD, Abbott2019PhRvD, Abbott2020PhRvD, Abbott2021PhRvD, LIGO2024arXiv}.

A massive star with an initial mass exceeding approximately $8\,M_\odot$ at the zero-age main sequence eventually reaches the final stage of its evolution when nuclear burning ceases and the thermonuclear energy sources are exhausted. During this phase, if the mass of the stellar core exceeds the effective Chandrasekhar mass limit, gravitational collapse becomes inevitable \citep{Baron_apj1990, Bethe_1990}. Several mechanisms have been proposed to explain how core-collapse supernovae (CCSNe) produce GWs, with the two most widely discussed being the neutrino-driven mechanism \citep{Janka_2012, Bethe_1990, Bethe_apj1985} and the magnetorotational mechanism \citep{Scheidegger_2008, Janka_2012, Kotake_2012PTEP, Mezzacappa_2014}. 
Some studies have also used ground-based GW detectors to constrain the GW emission from CCSNe \citep{Diao2025arXiv}. In the early stages of a CCSN event, GWs generated during the collapse and subsequent bounce of rapidly rotating iron cores have been extensively investigated under various equations of state, yielding a broad library of predicted GW waveforms \citep{Abdikamalov2014PhRvD, Dimmelmeier2008PhRvD, Richers2017PhRvD}. Numerical simulations indicate that the majority of GW emission occurs within approximately one second following core bounce, predominantly driven by convective motions and the standing accretion shock instability (SASI) \citep{Blondin_apj2003, Andresen_mn2017, Muller2012AA, Kuroda2016ApJL, Morozova2018ApJ, Yakunin_2017, Vartanyan2023PhysRevD, Wang2023ApJ}. In addition, significant progress has been made in the theoretical studies of SASI and in the methods for extracting and analyzing its signatures from GW data, providing valuable insights for future investigations of the dynamical processes in CCSN sources \citep{O'Connor2018ApJ, Takeda2021PhRvD, Powell2025arXiv}.

In an ideal scenario, disregarding transient noise and instrumental glitches \citep{Nuttall2015CQGra}, the reconstruction of GW signals CCSNe mainly depends on our understanding of their time-frequency structures \citep{Hayama_prd2015, Gossan2016PhRvD}. Since GWs generated by CCSNe are highly stochastic and significantly affected by factors such as the mass of the progenitor star and the equation of state (EoS) of the resulting proto-neutron star \citep{Muller2016PASA, Morozova2018ApJ}, matched filtering methods, commonly used in searches for compact binary coalescences, cannot be directly applied to CCSN GW searches \citep{Owen1999PhRvD, McIver2015}. For recent progress in this field, a comprehensive discussion can be found in \cite{Andresen2024arXiv}. In recent years, various methods have been proposed and applied to detect, reconstruct, and classify GWs from CCSNe. These include wavelet analysis (successfully employed in reconstructing GWs from compact binary mergers) \citep{McIver2015, Mezzacappa2024arXiv}, ensemble empirical mode decomposition (EEMD) \citep{Yuan2024MNRAS, Takeda2021PhRvD, Hu2022ApJ}, principal component analysis (PCA) \citep{Heng2009CQG, Rover2009PhRvd}, dynamic time warping (DTW) \citep{Suvorova2019PhRvD}, time-frequency analysis methods \citep{Yuan2025MNRAS}, and the increasingly popular machine learning techniques \citep{Chan2020PhRvD, Mitra2024MNRAS, Powell2024PhRvD}. However, most of these approaches have been developed and tested using data from aLIGO and aVirgo, and are typically effective only for reconstructing or classifying GW signals from sources within 100~kpc. More recently, \cite{Yuan2025MNRAS} extended the reconstruction range to approximately 300~kpc by utilizing the ET noise model of the third-generation ground-based GW detectors. Nevertheless, for the recent supernova event SN 2023ixf, no GW signal has yet been successfully extracted from the data \citep{LIGO2024arXiv}.

This underscores the need to explore more sensitive and robust analysis paradigms. Among emerging technologies, deep learning (DL), distinguished by its exceptional feature extraction capabilities, has become a vital complementary approach in GW data analysis. While models based on convolutional neural network (CNN) architectures have demonstrated significant potential in GW feature representation, existing supervised learning frameworks still face challenges in controlling the false alarm rate when high noise levels from distant sources. Given that backbone networks such as ResNet \citep{he2016deep} have matured in feature extraction, we suggest that the key to further enhancing model performance under low signal-to-noise ratio (SNR) and complex noise backgrounds lies not merely in increasing network depth, but in the optimization objective, specifically, the design of the Loss Function. This function directly dictates the distributional topology of signals and noise within the feature space.

Traditional classification tasks extensively employ cross-entropy (CE) as the loss function. By optimizing the posterior probability distribution to approximate discrete ground truth labels, this method performs exceptionally well in closed-set classification tasks. However, when applied to GW searches characterized by the unknown, its limitations regarding Feature Manifold construction become apparent:

\begin{enumerate}
\item \textbf{Lack of Intra-class Compactness and Generalization Bottlenecks:} CE loss primarily focuses on the delineation of decision boundaries (i.e., inter-class separability) without explicitly constraining the compactness of samples within the same class in the feature space. This limitation is well-documented in face recognition research\citep{wen2016discriminative}. Consequently, the features captured by the model may rely heavily on the local statistical properties of the training data, resulting in insufficient generalization robustness when encountering CCSNe signals with diverse morphologies and complex evolutionary paths.
\item \textbf{Sensitivity to Out-of-Distribution (OOD) Samples:} Classifiers trained with CE inherently rely on a "closed-world" assumption, presuming that all inputs belong to a predefined set of categories. However, real-world data streams from detectors contain a vast array of unknown, non-stationary noise morphologies (i.e., OOD samples). Under these conditions, models frequently misclassify unknown noise as known signals with high confidence \citep{hendrycks2016baseline}, posing a critical risk for GW astronomy, which demands an extremely low false alarm rate.
\end{enumerate}

To address these challenges and further exploit the potential of deep learning in weak signal identification, this paper introduces the contrastive learning paradigm \citep{le2020contrastive}. Unlike approaches that directly fit classification labels, contrastive learning aims to construct a structured, metric-defined Embedding Space\citep{wang2020understanding}. Its core logic lies in optimizing the relative distances of samples within the feature space through a metric learning process that "pulls positive pairs closer and pushes negative pairs apart." This paradigm delivers two significant benefits to the detection and classification of CCSNe GW signals:

First, by reinforcing intra-class compactness and maximizing inter-class separability, contrastive learning compels the model to prioritize feature separability, aiming to capture intrinsic signal morphology. This highly structured feature representation is poised to establish clearer classification boundaries under low SNR conditions, thereby enhancing the model's discriminative capability for complex waveforms. However, we also investigate how this optimization objective interacts with intrinsic dataset properties, such as amplitude distribution. Second, this framework provides a more natural geometric perspective for handling OOD samples\citep{seifi2024ood}. Within the feature space constructed by contrastive learning, authentic GW signals cluster tightly, whereas random noise or unknown glitches lacking physical signal characteristics tend to be projected into regions distant from these signal clusters. This distance-based topological separation enables the effective identification and rejection of OOD samples through the establishment of distance thresholds, significantly reducing the false alarm rate without compromising detection sensitivity. Therefore, the introduction of contrastive learning represents not merely an optimization of existing classification methods, but a substantial step towards building a high-reliability, low-false-positive GW data analysis pipeline.

The remainder of this paper is organized as follows: Section \ref{sec:method} introduces the proposed model architecture, with a focus on the mathematical construction of the contrastive learning framework and its loss function; Section \ref{sec:data_gen} details the simulation data generation process used for training and testing; Section \ref{sec:result} presents the classification performance evaluation and the topological analysis of the feature space; finally, Section \ref{sec:dis} provides a discussion of the results and prospects for future work.
\section{Methodology}
\label{sec:method}

\subsection{Model Architecture}
\label{sec:model}

\begin{figure}
    \centering
    \includegraphics[width=\linewidth]{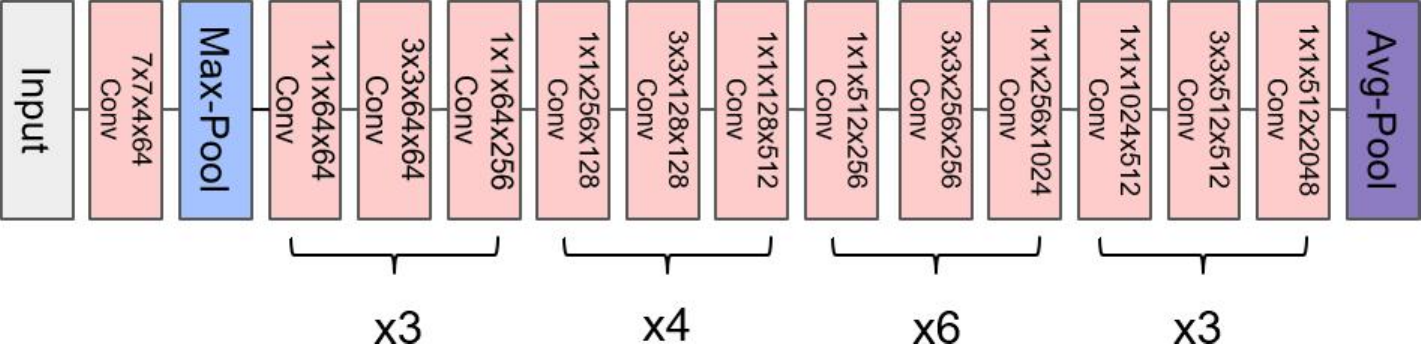}
    \caption{Schematic of the ResNet-50 encoder architecture. This model is adapted from the standard ResNet-50 classifier by removing the final classification head. The feature vector, output by the terminal global average pooling layer, is then utilized for downstream tasks. For visual clarity, residual connections between layers are omitted.}
    \label{fig:resnet50}
\end{figure}

We employ the standard ResNet-50 architecture\citep{he2016deep} as our detection model. The ResNet series of models is widely utilized across various computer vision tasks and has achieved state-of-the-art results. Furthermore, pre-trained ResNet models serve as the standard visual encoder in some vision-language large models. To accommodate our four-channel data, we modified the input channel dimension of the original model from 3 to 4. We also adjusted the number of classes in the output layer to 3, representing rotational signals, neutrino-driven signals, and a noise class, respectively. The model architecture is illustrated in Figure~\ref{fig:resnet50}.

\subsection{Contrastive Learning Training Pipeline}
\begin{figure}
    \centering
    \includegraphics[width=0.9\linewidth]{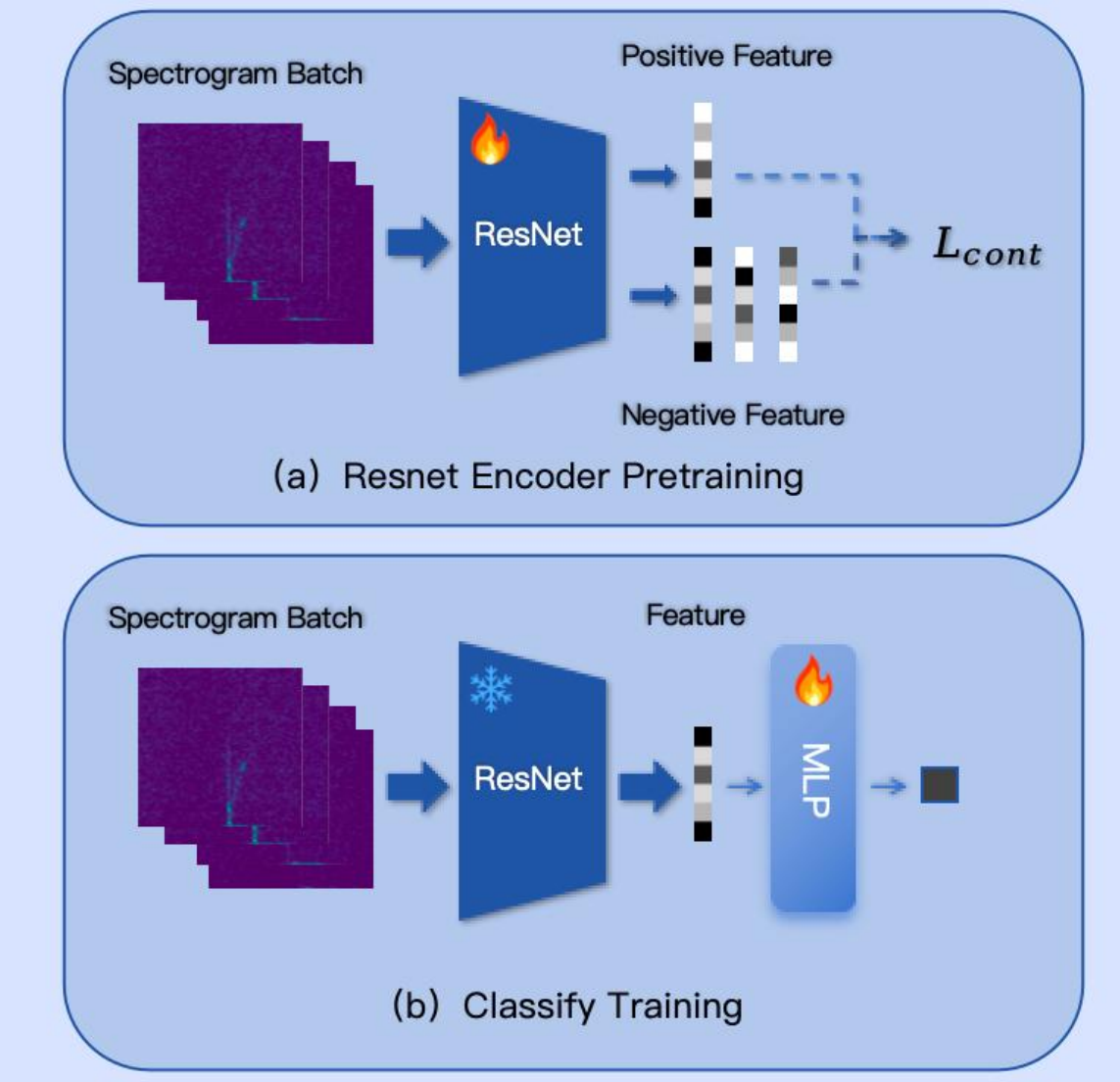}
    \caption{The two-stage training pipeline for the contrastive learning model. (a) The contrastive pre-training stage, where the ResNet encoder (indicated by the fire icon) is trained to optimize a contrastive loss ($L_{cont}$). (b) The downstream classification training stage, where the parameters of the ResNet encoder are frozen (indicated by the snowflake icon), and its output features are used to train a separate MLP classifier.}
    \label{fig:train_pipeline}
\end{figure}
Our training pipeline is divided into two distinct stages, as illustrated in Figure~\ref{fig:train_pipeline}. The first stage consists of supervised contrastive pre-training \citep{khosla2020supervised}. As shown in Figure~\ref{fig:train_pipeline}(a), each input signal is passed through a ResNet encoder to generate a feature vector. Signals from the same class are treated as positive pairs, while signals from different classes are treated as negative pairs. The contrastive loss is then calculated to train the ResNet encoder, enabling it to construct a well-structured feature space. 

Upon completion of the pre-training stage, we freeze the parameters of the ResNet encoder and append a three-layer MLP classifier, as depicted in Figure~\ref{fig:train_pipeline}(b). The objective of this second stage is to fine-tune the classifier to map the extracted feature vectors to their corresponding class labels. For this classification task, we employ the standard CE loss function.

\subsection{Contrastive Loss Function}

Our contrastive loss function is based on the Momentum Contrast (MoCo) \citep{he2020momentum} framework, which hinges on two core components: a dynamic dictionary queue and a momentum-updated encoder.

First, MoCo implements the dictionary as a dynamic First-In-First-Out (FIFO) queue. During each training step, the encoded features (''keys'') of the current mini-batch are enqueued, while the oldest batch of keys is dequeued. This mechanism allows for the maintenance of a large and consistent dictionary of negative samples, which significantly enhances the quality of contrastive learning.

Second, to ensure high consistency among the features within the queue, MoCo employs a momentum update strategy. It maintains two separate encoders: a \textit{query encoder} with parameters $\theta_q$, which is updated via standard backpropagation, and a \textit{momentum encoder} with parameters $\theta_k$. The momentum encoder is not updated directly by the loss gradient. Instead, its parameters are updated as a moving average of the query encoder's parameters:
\begin{equation}
\theta_k \leftarrow m\theta_k + (1-m)\theta_q
\end{equation}
where $m \in [0, 1]$ is the momentum coefficient. A high value of $m$ ensures that the momentum encoder evolves very slowly, thereby maintaining strong feature consistency among the keys in the dictionary.

The objective function is the InfoNCE (Noise Contrastive Estimation) loss \citep{oord2018representation}, defined as:
\begin{equation}
L_{q} = -\log \frac{\exp(q \cdot k_{+} / \tau)}{\exp(q \cdot k_{+} / \tau) + \sum_{i=0}^{K-1} \exp(q \cdot k_{i} / \tau)}
\end{equation}

In this formula, $q \cdot k$ denotes the dot product of the L2-normalized query and key features (i.e., their cosine similarity). The term $\tau$ is a temperature hyperparameter that modulates the sharpness of the softmax distribution.

\section{Data Generation}
\label{sec:data_gen}
To synthesize a high-fidelity data sample for training, we have designed a detailed data generation pipeline for the detector network. 
This pipeline aims to simulate how a GW signal is received by a global network of detectors (LIGO Hanford, LIGO Livingston, Virgo, and KAGRA) and ultimately transformed into a format suitable for deep learning model training.

\subsection{Coherent Waveform Projection and Time Delay}
Our simulation begins with a noiseless, source-frame CCSN theoretical GW waveform. The rotational CCSN signals are sourced from \citep{Abdikamalov2014PhRvD, Dimmelmeier2008PhRvD, Richers2017PhRvD}\footnote{\protect\url{https://stellarcollapse.org/gwcatalog.html}, \protect\url{https://zenodo.org/record/201145.}}, with the GW waveforms generated by \cite{Dimmelmeier2008PhRvD} being based on 2-D simulations. For the neutrino-driven mechanism, we employ waveforms from \citep{Mezzacappa2020PhRvD, Ott2013ApJ, Andresen2019MNRAS, Kuroda2017ApJ, Muller2012AA, Powell2019MNRAS, Radice2019ApJL, Powell2020MNRAS, Vartanyan2023PhysRevD}\footnote{\protect\url{https://wwwmpa.mpa-garching.mpg.de/ccsnarchive/data/Andresen2019/}, \protect\url{https://www.astro.princeton.edu/~burrows/gw.3d/}}. As shown in Table \ref{tab:waveforms}, these simulations cover a wide range of progenitor masses, from 9~$M_\odot$ to 60~$M_\odot$. Since these waveforms are generated under varying conditions (e.g., different distances, sampling rates, and durations), normalization is necessary before generating the time series. To achieve this, we scale the waveform amplitudes by relocating the sources to a standardized distance of 10~kpc from Earth. Additionally, to ensure uniform sampling rates, all waveforms are down-sampled to a preselected rate of 4096~Hz.

\begin{table*}
\centering
\caption{The mass ranges and mechanisms of the progenitors associated with the simulated waveforms used in this work. The ``\#'' column indicates the method ID number. The second column lists the corresponding study. The ``Mechanism'' column indicates the explosion mechanism for each waveform, with ``R'' representing the rotational mechanism and ``N'' indicating the neutrino-driven mechanism. The ``Mass'' column specifies the progenitor masses in solar mass units as represented in the simulations. The ``No.'' column denotes the number of waveforms available from each study. All listed masses correspond to the stellar masses at zero age unless otherwise noted.}\label{tab:data}
\label{tab:waveforms}
\begin{tabular}{lllll}
\hline
\hline
\# & & Mechanism & Mass ($M_{\odot}$) & No. \\ \hline
1 & Abdikamalov \citep{Abdikamalov2014PhRvD}  & R        & 12.0                          & 92   \\ 
2 & Dimmerlmeier \citep{Dimmelmeier2008PhRvD} & R        & 11.2, 15.0, 20.0, 40.0        & 136  \\ 
3 & Richers \citep{Richers2017PhRvD}      & R       & 12.0                          & 1824 \\ 
4 & Andresen \citep{Andresen2019MNRAS}     & N        & 11.2, 15.0                    & 6    \\ 
5 & Muller \citep{Muller2012AA}       & N        & 15.0, 20.0                    & 6    \\ 
6 & Ott \citep{Ott2013ApJ}          & N        & 27                            & 8    \\ 
7 & Powell \citep{Powell2019MNRAS}       & N        & 18.0                     & 1    \\ 
8 & Powell \citep{Powell2020MNRAS}       & N       &  39.0                     & 1    \\ 
9 & Radice \citep{Radice2019ApJL}       & N        & 9, 10, 11, 12, 13, 19, 25, 60 & 8    \\ 
10 & Vartanyan \citep{Vartanyan2023PhysRevD}  &N    & 9, 9.25, 9.5, 11, 12.25, 14, 15.01, 23 & 11   \\
11 & Mezzacappa \citep{Mezzacappa2020PhRvD}      & N       & 15                            & 1    \\ 
12 & Kuroda \citep{Kuroda2017ApJ}       & N        & 11.2, 15.0                    & 2    \\
\hline
\hline\\

\end{tabular}

\end{table*}
This theoretical waveform includes two orthogonal polarization components: the plus-polarized strain $h_+(t)$ and the cross-polarized strain $h_\times(t)$. To project this theoretical waveform onto the various detectors on Earth, we first randomly sample the source's sky location in a celestial coordinate system, defined by the right ascension $\alpha \in [0, 2\pi]$ and the declination $\delta \in [-\pi/2, \pi/2]$.For the $i$-th detector, its response to the GW is determined by the antenna pattern functions, $F_{+,i}$ and $F_{\times,i}$. These two functions are dependent on the source's sky location $(\alpha, \delta)$. Therefore, the strain signal $h_i(t)$ recorded at the $i$-th detector can be expressed as:
\begin{equation}
h_i(t) = F_{+,i}(\alpha, \delta)h_+(t_i) + F_{\times,i}(\alpha, \delta)h_\times(t_i)
\end{equation}

Here, $t_i$ is the time at which the signal arrives at the $i$-th detector. Due to the propagation of GWs at the speed of light, $c$, detectors at different spatial locations will receive the signal at different times, introducing a time delay $\Delta t_i$. Taking the geocenter as the reference point, this delay can be precisely calculated as:
\begin{equation}
\Delta t_i = \frac{\vec{r}_i \cdot \vec{\Omega}}{c}
\end{equation}

where $\vec{r}_i$ is the vector from the geocenter to the $i$-th detector, and $\vec{\Omega}$ is the unit vector from the geocenter towards the direction of the GW source, which is uniquely determined by $(\alpha, \delta)$. We then apply this time delay $\Delta t_i$ to the signal $h_i(t)$ of each channel, thereby generating multiple channels of noiseless signals that maintain a coherent temporal relationship.
\subsection{ Noise Injection and Signal Scaling}

The actual signal from a detector is a superposition of the GW signal $h_i(t)$ and the detector's intrinsic noise. To simulate this, we use the power spectral density (PSD), denoted as $S_{n,i}(f)$, for LIGO Hanford, LIGO Livingston, Virgo, and KAGRA at design sensitivity\citep{abbott2020prospects}. Using the Bilby library \citep{ashton2019bilby}, we then generate noise $n_i(t)$ that is statistically consistent with the detector's characteristics based on this PSD.

Subsequently, this noiseless signal $h_i(t)$ is injected into the simulated noise background. At the same time, to account for GW events at different distances, we introduce a scaling factor $\lambda$. We assume our baseline waveform $h_i(t)$ corresponds to a source at a reference distance $D_{\text{ref}} = 10~\text{ kpc}$. The signal amplitude from a source at a distance $D$ will be inversely proportional to the distance. Therefore, the scaling factor $\lambda$ is defined as:
\begin{equation}
\lambda = \frac{D_{\text{ref}}}{D}
\end{equation}
We perform random sampling for the distance $D$ within the time interval $[10~\text{ kpc}, 200~\text{ kpc}]$, resulting in a 10-second signal. Ultimately, the final time-domain signal $s_i(t)$ at the detector is expressed as:
\begin{equation}
s_i(t) = \lambda \cdot h_i(t) + n_i(t)
\end{equation}

\subsection{Time-Frequency Transformation and Data Formatting}

After obtaining the time-series signal $s_i(t)$, we first perform a whitening process. This step transforms the colored Gaussian noise into white Gaussian noise, effectively suppressing frequency bins where the noise power is high.
Next, we transform the whitened time-series signal into a time-frequency representation (Spectrogram) using the short-time fourier transform (STFT). In this process, we set the sampling rate to $f_s = 4096~\text{ Hz}$, the STFT window length to 510 samples, and the overlap between windows to 490 samples.

We enhance the dataset by standardizing the input dimensions of the long-duration background noise spectrograms, which are characteristic of CCSNe GW signals. Specifically, we crop the spectrograms to a size of $256 \times 1024$ pixels, where 256 corresponds to the frequency resolution dimension (preserving the complete frequency range), and 1024 corresponds to the time-step dimension (approximately 10 seconds). During the cropping process, we apply a random positioning algorithm, which calculates the random position of the energy peak's occurrence within this 10-second window. This greatly enhances the model's invariance to the signal's appearance time.

Through the pipeline described above, we ultimately generate a four-channel spectrogram for each GW event with a shape of $(4, 256, 1024)$. This includes the whitened data from the four detectors, along with the sky location and distance information of the simulated source, which are finally processed into a unified tabular representation of the GW signal. These structures constitute the training set required for model training and evaluation.

\section{Results}
\label{sec:result}
\subsection{Feature Visualization}
\begin{figure}
    \centering
    \includegraphics[width=0.9\linewidth]{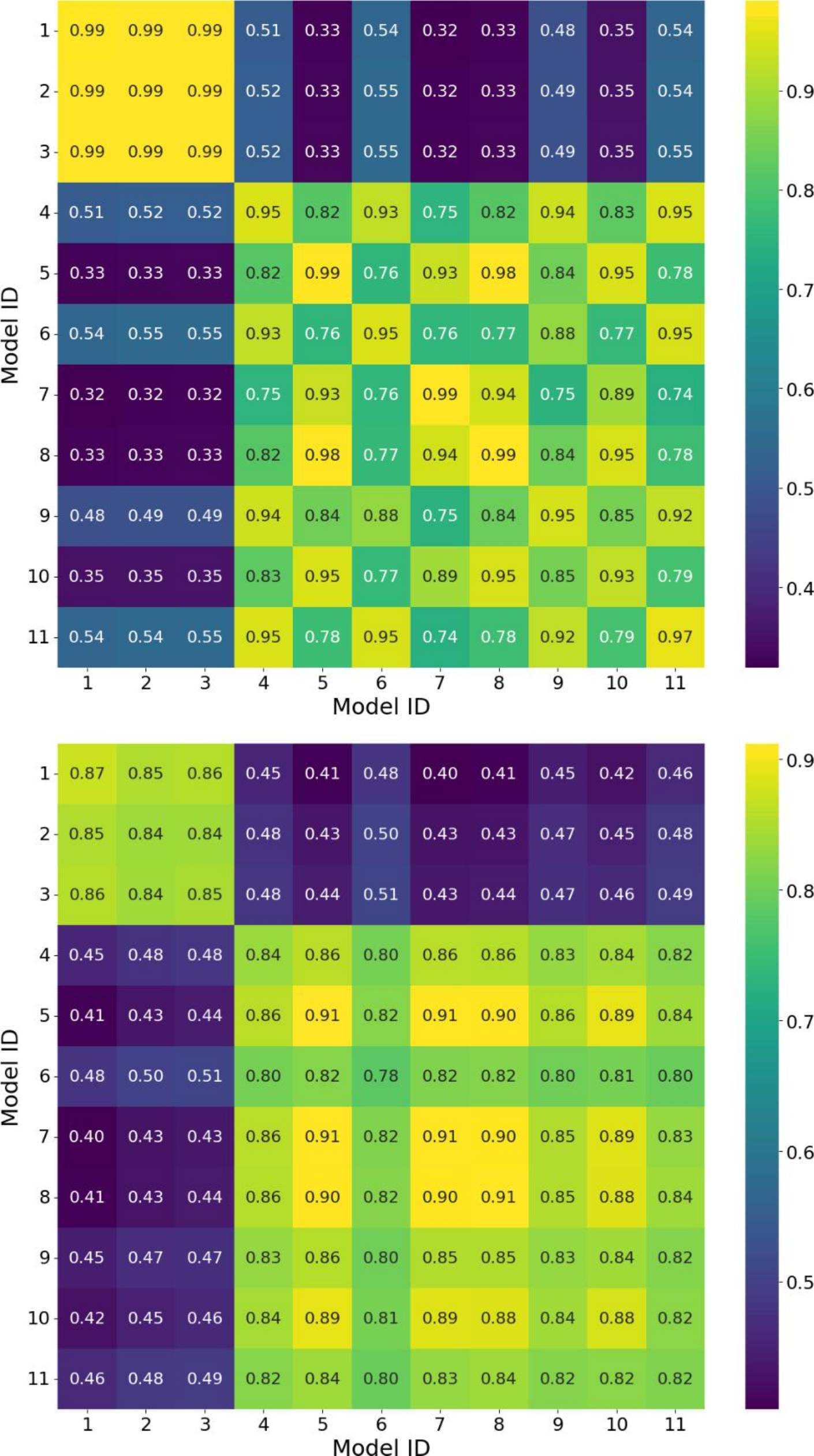}
    \caption{Feature Similarity Matrices. The horizontal and vertical axes represent the GW model numbers, corresponding to the model names in Table 1. The top panel displays the cosine similarity matrix derived from features learned through our contrastive learning method, while the bottom panel presents the matrix from a directly end-to-end trained model. Notably, features produced by the contrastive learning algorithm demonstrate superior intra-class similarity.}
    \label{fig:feature_sim}
\end{figure}

\begin{figure}
    \centering
    \includegraphics[width=0.9\linewidth]{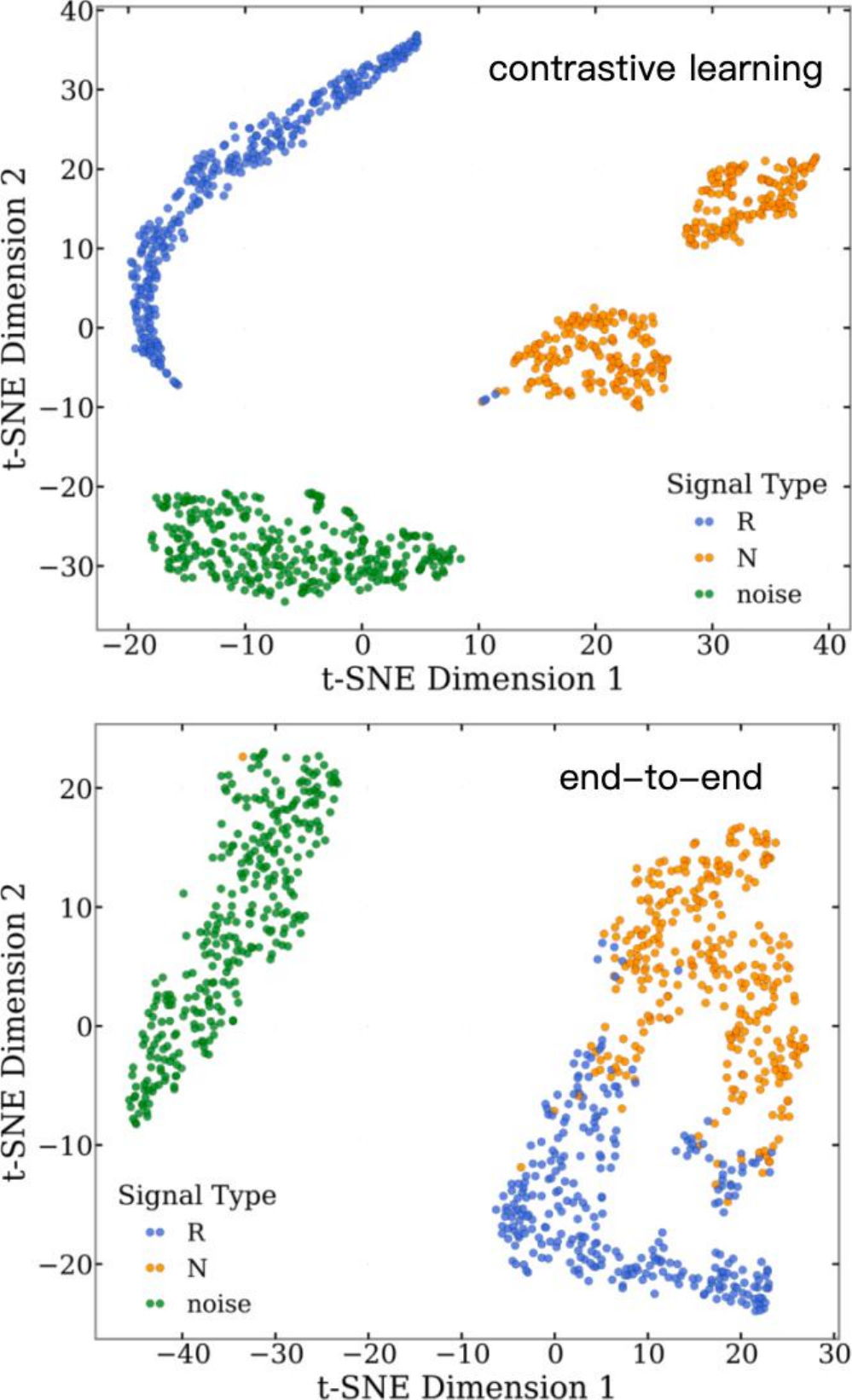}
    \caption{Comparison of t-SNE Feature Space Visualizations. The top panel illustrates the feature distribution derived from our contrastive learning method, while the bottom panel corresponds to features from a model trained end-to-end. The features produced by contrastive learning are clearly distributed more compactly.}
    \label{fig:t-SNE}
\end{figure}

Following training via contrastive learning, our ResNet model effectively encodes four-channel spectrograms into feature vectors that encapsulate salient class information. To visualize the separability of these embeddings, we computed the pairwise feature cosine similarity matrix, presented in Figure~\ref{fig:feature_sim}. This matrix is constructed by calculating the similarity $\mathbf{S}_{ij} = (\mathbf{v}_i \cdot \mathbf{v}_j) / (\|\mathbf{v}_i\| \|\mathbf{v}_j\|)$ for all pairs of feature vectors $(\mathbf{v}_i, \mathbf{v}_j)$ in the test set. The matrix exhibits a pronounced block-diagonal structure, which signifies high intra-class similarity and low inter-class similarity, demonstrating that our approach successfully clusters the rotational and neutrino-driven GW signals in the learned embedding space. In contrast, the embedding space generated by a standard, end-to-end trained ResNet-50 classification model appears significantly more diffuse. Furthermore, we employed t-distributed stochastic neighbor embedding (t-SNE), a non-linear dimensionality reduction technique that visualizes high-dimensional data by preserving local neighborhood structures in a low-dimensional map. We randomly sampled 900 test signals with distances ranging from 10-200~kpc to generate the t-SNE plot, shown in Figure~\ref{fig:t-SNE}. The plot for our contrastive learning model clearly shows that the rotational and neutrino-driven signal clusters are substantially more compact (high intra-class cohesion) and maintain a greater spatial distance from one another (low inter-class coupling). Conversely, the results from the end-to-end trained model yield more scattered clusters, with a noticeable "mutual infiltration" or overlap between the rotational and neutrino-driven class features in the embedding space. This pronounced class ambiguity is evidently detrimental to robust signal classification.

\subsection{Classification Results}
To evaluate the classification efficacy of our model, we generated a comprehensive synthetic test set comprising 350,000 samples. This dataset includes 50,000 pure noise instances, as well as 150,000 neutrino-driven signals and 150,000 rotational signals. Events were simulated at three distinct source distances: 10, 200, and 1200~kpc, with 50,000 samples at each distance. The 10~kpc and 200~kpc distances represent the two distance extremes included in the training set, while the 1200~kpc data are used to explore whether the model can extrapolate to greater distances.

We plotted the receiver operating characteristic (ROC) curves comparing the contrastive learning algorithm and the end-to-end approach at different distances for both rotational and neutrino-driven signals. As illustrated in the figures, the ROC curve is a commonly used method for evaluating classifier performance, displaying the true positive rate (TPR) at various false positive rate (FPR) thresholds. For our three-class model, the FPR is defined as the proportion of samples from the other two classes that are incorrectly classified as the true class depicted in the ROC curve, while the TPR represents the proportion of true class samples that are successfully detected. A higher TPR at a given FPR indicates stronger model performance.

\begin{figure}
    \centering
    \includegraphics[width=0.9\linewidth]{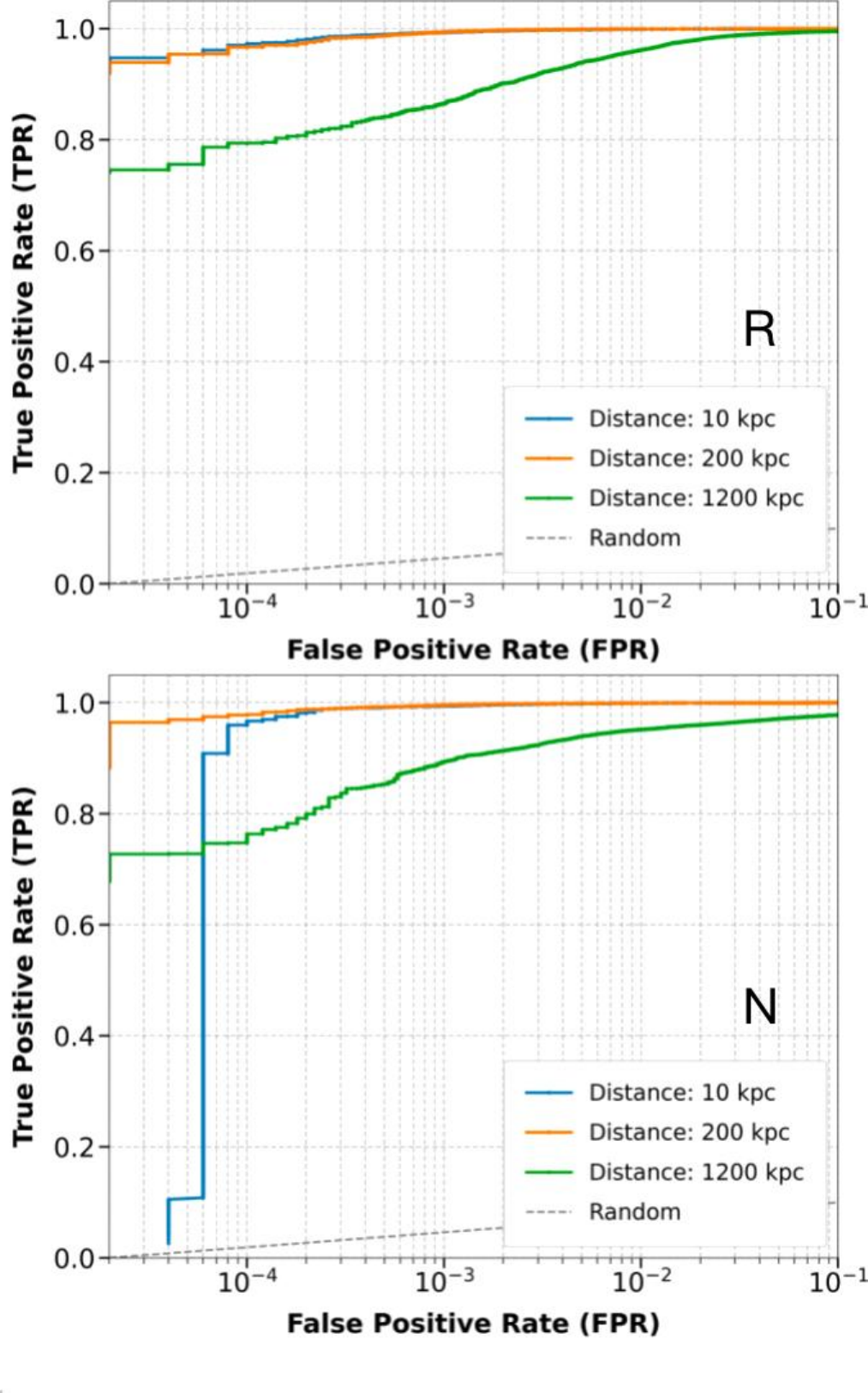}
    \caption{ROC curves for the contrastive learning model. The upper panel shows the classification performance for rotational signals, while the lower panel shows neutrino-driven signals. The TPR is evaluated against the FPR on a logarithmic scale at varying source distances: 10~kpc (blue), 200~kpc (orange), and 1200~kpc (green). The dashed line represents the baseline performance of a random classifier (TPR=FPR).}
    \label{fig:ROC_cont}
\end{figure}

\begin{figure}
    \centering
    \includegraphics[width=0.9\linewidth]{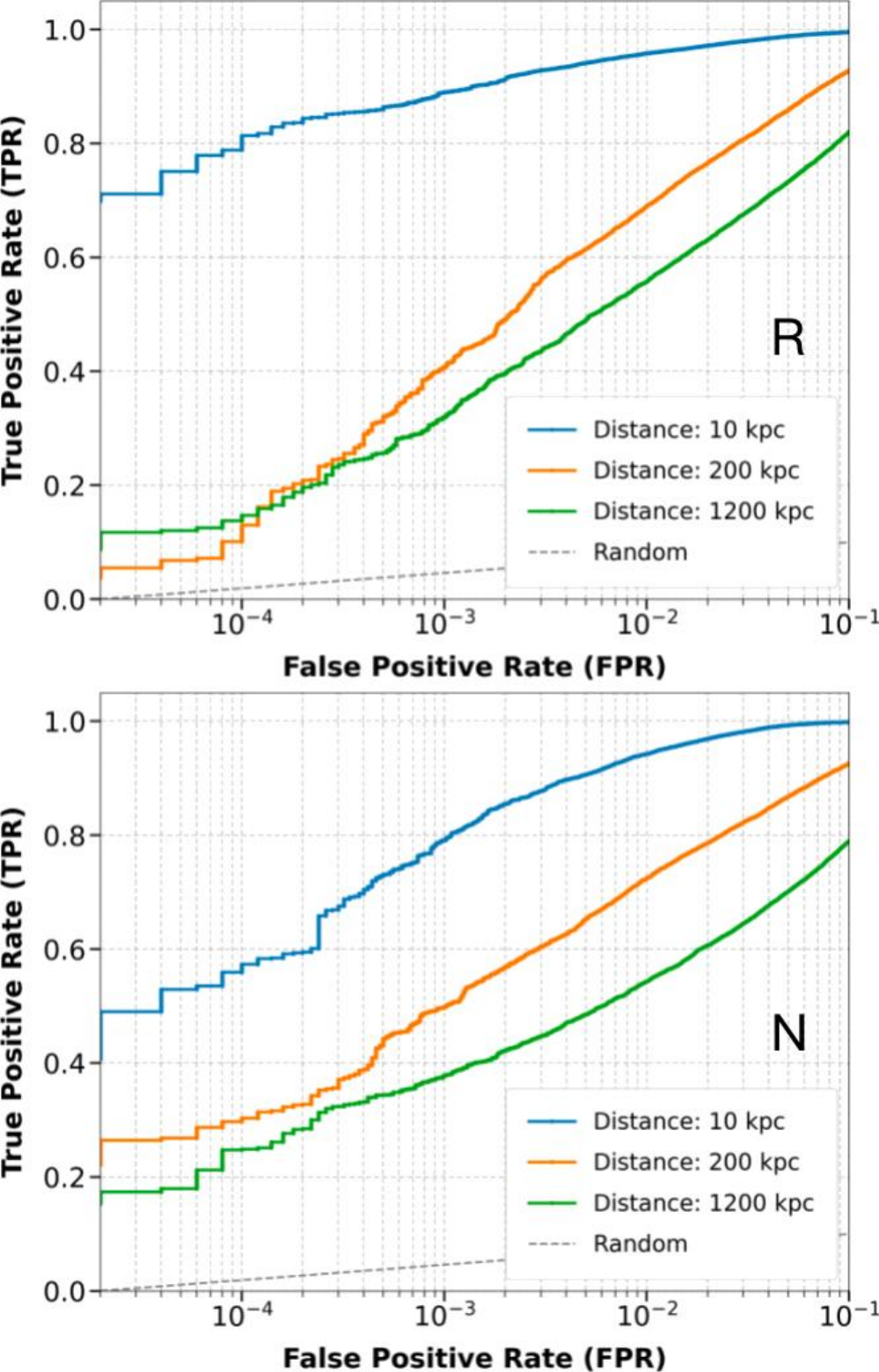}
    \caption{ROC curves for the end-to-end training model. The panels are consistent with Figure~\ref{fig:ROC_cont}.}
    \label{fig:ROC_e2e}
\end{figure}

The upper panels of Figure~\ref{fig:ROC_cont} and Figure~\ref{fig:ROC_e2e} display the ROC curves for classifying rotational signals using the contrastive learning model and the end-to-end model, respectively. It can be observed that the contrastive model achieves a near-ideal detection characteristic. For sources at 10 and 200~kpc (blue and orange curves), the TPR approaches nearly 100\% even at an extremely low FPR of $10^{-4}$, and even when extended to 1200~kpc, the TPR remains around 80\%. In contrast, when the FPR is controlled at $10^{-4}$, the end-to-end approach achieves only approximately 80\% TPR at a distance of 10~kpc, while the TPR rapidly decays to below 20\% for both 200~kpc and 1200~kpc. The lower panels of Figure~\ref{fig:ROC_cont} and Figure~\ref{fig:ROC_e2e} show the ROC curves for neutrino-driven signals using both methods. It can be seen that when the FPR exceeds $6 \times 10^{-5}$, the contrastive learning algorithm maintains performance similar to that observed for rotational signals. For the end-to-end algorithm, however, performance degrades at 10~kpc, achieving less than 60\% TPR at an FPR of $10^{-4}$, although performance slightly improves at 200~kpc and 1200~kpc, exceeding 20\% TPR. This is likely because neutrino-driven signals are generally weaker in intensity than rotational signals, causing the end-to-end model to be more inclined to classify weaker signals as neutrino-driven.

\begin{figure}
    \centering
    \includegraphics[width=\linewidth]{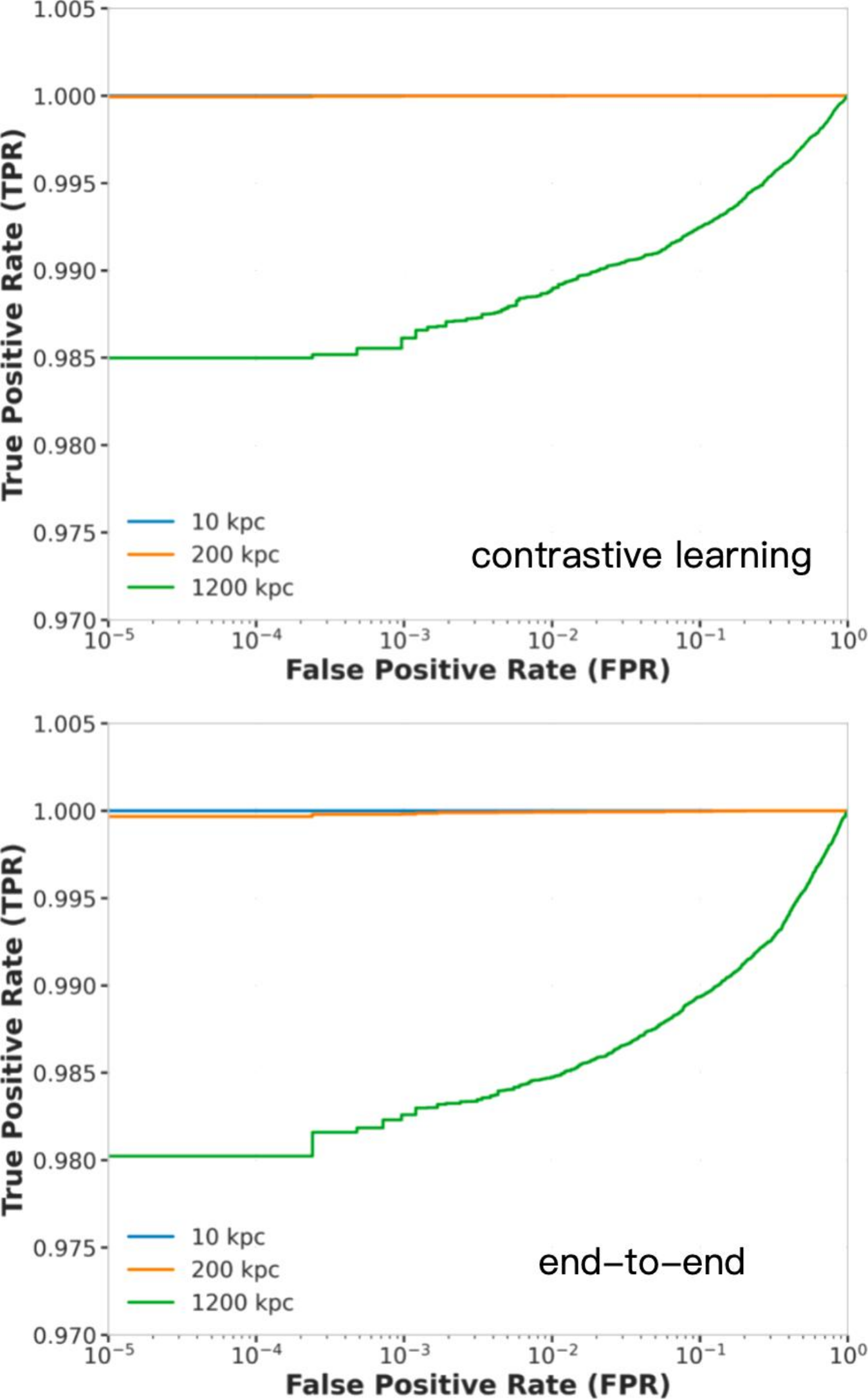}
    \caption{Comparison of ROC curves for signal detection performance. (Top) The contrastive learning algorithm. (Bottom) The end-to-end training algorithm. For this analysis, rotational and neutrino-driven signal types were not differentiated, focusing solely on the model's ability to detect signals against noise.}
    \label{fig:ROC_SN}
\end{figure}

It is noteworthy, as shown in the lower panel of Figure~\ref{fig:ROC_cont}, that the performance of the contrastive learning algorithm degrades sharply for neutrino-driven signals at specific distances when an FPR below $6 \times 10^{-5}$ is required. This degradation reveals a fundamental characteristic of the contrastive embedding space: its hypersensitivity to boundary outliers. Because the contrastive objective enforces extreme cluster compactness, "outlier" samples that lie in the morphological transition zone are not merely classified with low confidence (as in CE) but are often pulled decisively into the incorrect cluster. Consequently, these specific anomalous signals become difficult to filter out via simple threshold adjustment. This "over-confidence" on hard samples is a possible reason for this phenomenon, representing a trade-off for the superior separation of bulk samples.

Furthermore, we assessed the generalized signal detection capability, namely, discriminating any GW event (neutrino-driven or rotational) from pure noise, as shown in Figure~\ref{fig:ROC_SN}. Since the noise is Gaussian, signal detection by the model requires only capturing the deviation of the input signal distribution from a Gaussian distribution. Consequently, both methods achieved good performance on the generalized signal detection task. However, the contrastive learning model still outperforms the end-to-end model, achieving a TPR that is 0.5\% higher at 1200kpc when the FPR is controlled at $10^{-5}$.

\subsection{Waveforms Excluded from Training}
\begin{figure}
    \centering
    \includegraphics[width=\linewidth]{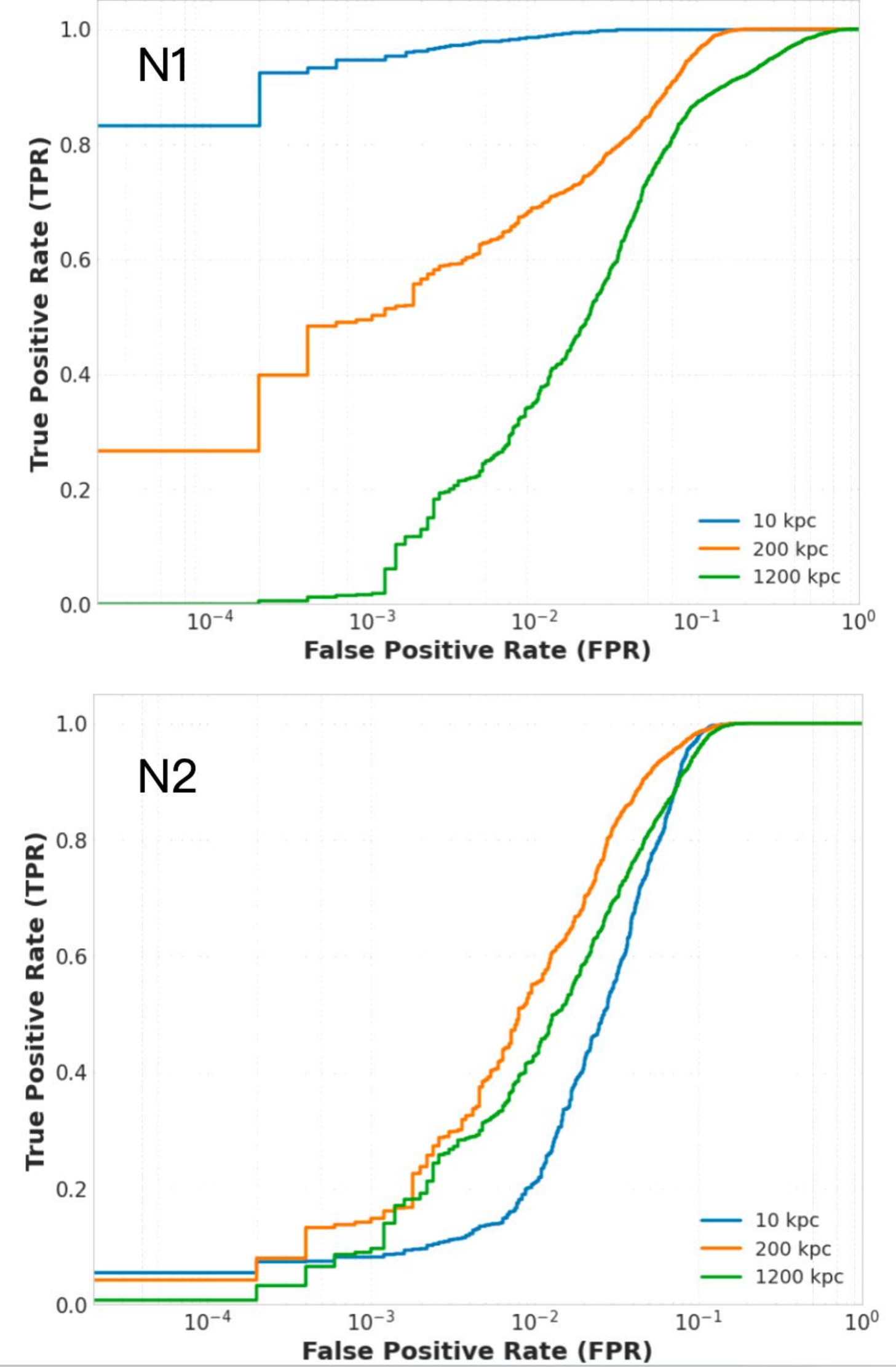}
    \caption{ROC curves for the excluded samples. (top) Performance for the N1 signal. (bottom) Performance for the N2 signal, which is an outlier characterized by an intensity far exceeding the majority of signals.}
    \label{fig:ROC_OOD}
\end{figure}

\begin{figure}
    \centering
    \includegraphics[width=\linewidth]{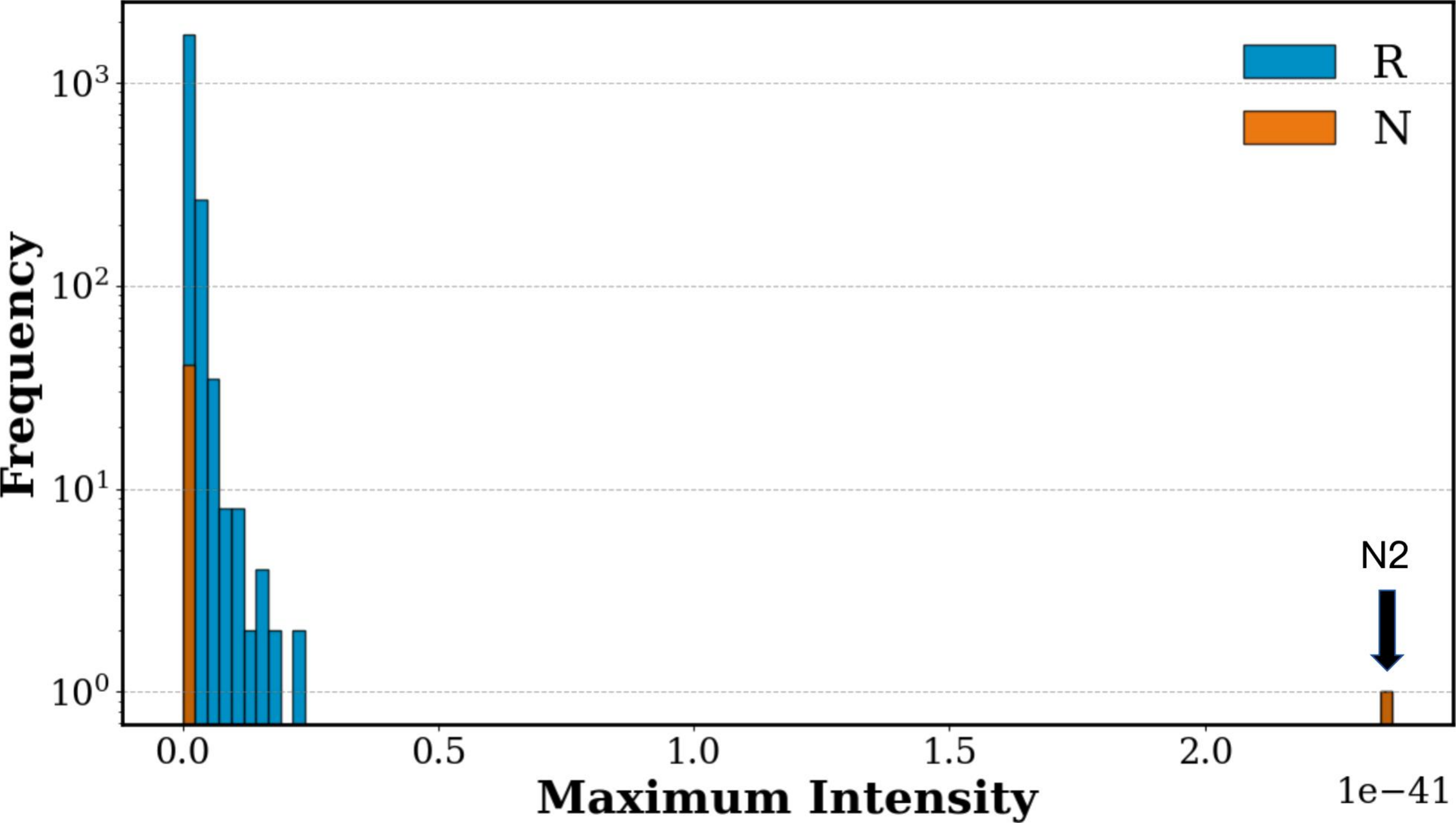}
    \caption{Distribution of maximum signal power at 10~kpc for rotational and neutrino-driven signals. The histogram reveals a distinct amplitude bias, showing that rotational signals are generally stronger than neutrino-driven signals. The anomalous outlier signal visible on the far right of the plot corresponds to N2, which is a neutrino-driven signal.}
    \label{fig:signal_distribute}
\end{figure}

To evaluate the model's detection and classification performance on previously unseen GW signals, we withheld the waveforms from \citet{Kuroda2017ApJ} from the training set. These signals belong to the neutrino-driven category, comprising two waveforms: S11.2\_GW\_Nu (N1) and S15.0\_GW\_Nu (N2). Notably, N2, despite being a neutrino-driven signal, possesses a peak amplitude that significantly surpasses that of typical rotational signals, as shown in Figure~\ref{fig:signal_distribute}. To test the classification performance of the contrastive learning model on N1 and N2 signals, we randomly generated 75,000 samples for each signal, with 25,000 samples at each of the three distances: 10~kpc, 200~kpc, and 1200~kpc. Additionally, we sampled 25,000 noise instances and 25,000 rotational signals at each of the three distances as negative samples. Using the same methodology as described in Section~\ref{sec:result}, we plotted the ROC curves as shown in Figure~\ref{fig:ROC_OOD}. For the N1 signal (which exhibits a nominal amplitude), the model maintains robust classification performance within the 10--200~kpc range.  Although significant performance degradation is observed at 1200~kpc, where detection becomes essentially impossible at an FPR of $10^{-4}$, the model still achieves over 80\% TPR at 10~kpc and more than 20\% TPR even at 200~kpc. Conversely, for the N2 signal with its anomalous amplitude, the model performs poorly; at 10~kpc, it is more likely to be misclassified as a rotational signal, achieving less than 20\% TPR even at an FPR of 1\%. The performance at 10~kpc is even worse than at 200~kpc and 1200~kpc, demonstrating that even the contrastive learning algorithm tends to classify higher-intensity signals as rotational signals. This indicates that, in the absence of explicit normalization, the contrastive objective exploits the most discriminative global parameter, i.e., amplitude, to maximize cluster separation. We visualized the amplitude distributions for both rotational and neutrino-driven signal types, as shown in Figure~\ref{fig:signal_distribute}. It is evident that rotational signals are, on average, significantly stronger than neutrino-driven signals. This result underscores that ``morphological learning'' cannot be guaranteed by the loss function alone, as the model may instead leverage intensity differences as a discriminative feature.


\section{Discussion}
\label{sec:dis}

In this paper, we proposed a two-stage training strategy based on supervised contrastive learning for the classification of CCSN GW signals. This approach was chosen to overcome two primary challenges faced by the traditional CE loss in handling high-noise GW data: the lack of a fine-grained metric structure in the feature space and vulnerability to OOD samples. Our core hypothesis was that pre-training an encoder via contrastive learning to construct an embedding space with high intra-class compactness and strong inter-class separability would provide more robust and discriminative features for the downstream classification task. Our experimental results largely validate this hypothesis. Feature visualizations (Figures \ref{fig:feature_sim} and \ref{fig:t-SNE}) clearly demonstrate the significant advantage of the contrastive learning method in building a structured feature space. Compared to the diffuse, overlapping feature clusters produced by a standard end-to-end model, our method successfully separates and compresses rotational and neutrino-driven signals into two well-defined clusters in the embedding space. This superior feature representation translated directly into stronger classification performance. The ROC curves (Figures \ref{fig:ROC_cont}, \ref{fig:ROC_e2e}, and \ref{fig:ROC_SN}) show that our model outperforms the end-to-end baseline in distinguishing between rotational and neutrino-driven signals at various distances (10~kpc, 200~kpc, and 1200~kpc). Notably, in the broader and more practical ``signal vs. noise'' detection task (Figure \ref{fig:ROC_SN}), the contrastive learning model achieved a significantly higher TPR at the same FPR, highlighting its potential for extracting real signals from noisy backgrounds.

However, our analysis also revealed important limitations and phenomena requiring further investigation. First, we observed a sharp degradation in the classification performance for neutrino-driven signals at extremely low FPRs ($< 6 \times 10^{-5}$) (lower panel of Figure \ref{fig:ROC_cont}). This appears to be a side effect of the "high-confidence" nature of contrastive learning: because the model strives to create highly separated and compact clusters, a misplaced sample can become deeply embedded within the wrong cluster, leading to a high-confidence error that is difficult to correct by adjusting the decision threshold. Second, the OOD test on the waveforms from \citet{Kuroda2017ApJ} (Figure \ref{fig:ROC_OOD}), which were excluded from training, exposed a deeper issue: the model's sensitivity to inherent dataset biases. The model performed poorly on the neutrino-driven signal (N2) with an anomalously high amplitude, misclassifying it as rotational. The amplitude distribution shown in Figure \ref{fig:signal_distribute} confirms a clear amplitude bias in our training set: rotational signals are, on average, significantly stronger than neutrino-driven signals. Consequently, the contrastive objective likely leveraged this "high amplitude" as a shortcut for maximizing separability. This suggests that while contrastive learning is highly effective for signal detection (where signal vs. noise separation is primary), utilizing it for fine-grained source classification might benefit from balancing the dataset intensity as a potential solution to ensure the model focuses on morphological features rather than intensity priors.

These findings provide important guidance for future work. While contrastive learning excels at building feature spaces, we must address the model's reliance on "spurious correlations" such as signal amplitude. We must also introduce techniques, similar to temperature scaling, to explore how to prevent the model from generating over-confident results, thereby addressing the performance degradation caused by a few anomalous samples. Furthermore, we can test the model's classification capability on OOD noise within real noise backgrounds to determine if the contrastive learning algorithm can effectively reduce false alarms on OOD noise, further exploring the potential for deep learning methods to be applied in real-world scenarios.

\section*{Data Availability}
The simulated CCSN GW spectrogram dataset generated and used in this study is available from the corresponding author upon reasonable request. The original rotational mechanism CCSN waveforms are available at \protect\url{https://stellarcollapse.org/gwcatalog.html} and \protect\url{https://zenodo.org/record/201145}. The neutrino-driven mechanism CCSN waveforms are available at \protect\url{https://wwwmpa.mpa-garching.mpg.de/ccsnarchive/data/Andresen2019/} and \protect\url{https://www.astro.princeton.edu/~burrows/gw.3d/}.

\bibliographystyle{apsrev}
\bibliography{apssamp}

\end{document}